\documentclass[aps,reprint,amsmath,amssymb,superscriptaddress,longbibliography,nofootinbib,showkeys]{revtex4-1}
\pdfoutput=1
\usepackage{graphicx,xcolor}
\usepackage{slashed}
\usepackage{float}
\usepackage[
    colorlinks,
    linkcolor=green!30!blue,           
    citecolor=red,            
    filecolor=green!30!blue,        
    urlcolor=green!30!blue,         
    hyperfootnotes]{hyperref}
\usepackage{bm}
\usepackage{url}
\usepackage{accents}
\usepackage[paperwidth=215.9mm,paperheight=279.4mm,centering,hmargin=2.1cm,vmargin=2cm]{geometry}

\usepackage{tcolorbox}
\usepackage{hyperref}
\usepackage{multirow}
\usepackage{tabularx}
\usepackage{array}
\usepackage{colortbl}
\tcbuselibrary{skins}
\newcolumntype{Y}{>{\raggedleft\arraybackslash}X}
\tcbset{tab1/.style={fonttitle=\bfseries\large,fontupper=\normalsize\sffamily,
colback=white,colframe=black,colbacktitle=Salmon!40!white,
coltitle=black,center title,freelance,frame code={
\foreach \n in {north east,north west,south east,south west}
{\path [fill=red!75!black] (interior.\n) circle (3mm); };},}}

\tcbset{tab2/.style={enhanced,fonttitle=\bfseries,fontupper=\normalsize\sffamily,
colback=white,colframe=black,coltitle=black,center title}}

\begin{document}
\title{{Limits on the Charged Higgs Parameters in the Two Higgs Doublet Model 
using CMS $\sqrt{s}=13$ TeV Results  }}

\author{Prasenjit Sanyal}
\email{psanyal@iitk.ac.in}
\affiliation{Department of Physics, Indian Institute of Technology Kanpur, Kanpur 208 016, India}


\begin{abstract}
 Latest CMS results on the upper limits on $\sigma_{H^\pm}$BR($H^\pm \rightarrow \tau^\pm\nu)$ and  $\sigma_{H^\pm}$BR($H^+ \rightarrow t\bar{b}$) for $\sqrt{s}=13$ TeV at an integrated luminosity of 35.9 fb$^{-1}$ are used to impose constraints on the charged Higgs $H^\pm$ parameters within the Two Higgs Doublet Model (2HDM). The 2HDM is the simplest extension of Standard Model (SM) under the same gauge symmetry to contain charged Higgs and is relatively less constrained compared to Minimal Supersymmetric Standard Model (MSSM). The latest results lead to much more stringent constraints on charged Higgs parameter space in comparison to the earlier 8 TeV results. The CMS collaboration also studied the exotic bosonic decay $H^\pm \rightarrow W^\pm A$ and $A \rightarrow \mu^+ \mu^-$ for the first time and put upper limits on the BR($t\rightarrow H^+ b$) for light charged Higgs. These constraints lead to exclusion of parameter space which are not excluded by the $\tau \nu$ channel. For comparison the exclusion regions from flavour physics constraints are also discussed.
\end{abstract}
\date{\today}
\maketitle
\section{Introduction}

 The Standard Model (SM) of particle physics is the most successful model in explaining nearly all particle physics phenomenology. The discovery of neutral scalar of mass 125 GeV with properties similar to the Higgs boson in SM \cite{Aad:2012tfa,Aad:2013wqa,Chatrchyan:2012xdj,Chatrchyan:2013lba} makes SM as the most acceptable model of particle physics. Despite being successful, the SM fails to explain the existence of dark matter, neutrino oscillation and matter-antimatter asymmetry. SM also does not explain the mass hierarchy in elementary particles and gravity is not included. Apart from that there is no fundamental reason to have only one Higgs doublet 
(i.e. minimal under the SM gauge symmetry) and the discovery of another scalar boson (neutral or charged) would require an extension of the SM. The simplest extension of SM under the same (SM) gauge symmetry is the Two Higgs Doublet Model (2HDM) \cite{Gunion:2002zf,Gunion:1989we,Branco:2011iw,Davidson:2005cw,Pich:2009sp,Bernon:2015qea,Bernon:2015wef}. So far there is no evidence for any other scalar upto a mass of few TeV and hence the parameter space of 2HDM is getting significantly constrained from experimental observations \cite{Aoki:2009ha,Mahmoudi:2009zx,Maniatis:2009by,Jung:2010ik,Chen:2013kt,Chiang:2013ixa,Coleppa:2013dya,Chen:2013rba,Bechtle:2015pma,Keus:2015hva,Akeroyd:2016ymd,Cacciapaglia:2016tlr,Krawczyk:2017sug,Arbey:2017gmh,Arhrib:2018ewj,Bhatia:2017ttp,Gori:2017qwg,Arhrib:2017wmo,BARAK2016896}. The scalar sector of 2HDM consists of five scalars, two $CP$ even scalars ($h$ and $H$), one $CP$ odd scalar (or pseudoscalar) $A$ and two charged Higgs $H^{\pm}$. The most general Yukawa sector (Type III) of 2HDM leads to flavour changing neutral currents (FCNC) at tree level. To avoid the FCNC, Glashow and Weinberg implemented a discrete symmetry in the Yukawa sector which leads to the four possible types of Yukawa interactions in 2HDM i.e. Type I, Type II, Type X (lepton specific) and Type Y (flipped model). A brief review on 2HDM is given in section  \ref{sec:2HDM review}. 
 
 The production of charged Higgs, depending on its mass with respect to the top quark, can be divided into light ($M_{H^\pm}\ll M_t$), intermediate ($M_{H^\pm}\sim M_t$) and heavy ($M_{H^\pm}\gg M_t$) scenarios \cite{Degrande:2016hyf,Flechl:2014wfa,Degrande:2015vpa,deFlorian:2016spz}. Throughout the analysis the alignment limit is considered i.e. $\sin(\beta-\alpha)\rightarrow 1$, (where the mixing angles $\beta$ and $\alpha$ are defined in section \ref{sec:2HDM review}) so that the neutral scalar $h$ behaves like the SM Higgs boson. The precisely measured electroweak parameter $T$ is highly sensitive on the mass splitting of $H^\pm, H$ and $A$. The alignment limit and minimum mass splitting restricts the charged Higgs decay mostly into the fermionic sector and the experimental constraints put exclusion bound on the charged Higgs ($M_{H^\pm}-\tan\beta$) parameter space. In this paper the 13 TeV CMS results \cite{Sirunyan:2019hkq,CMS:1900zym,Sirunyan:2019zdq} at an integrated luminosity of 35.9 fb$^{-1}$ are used to restrict the charged Higgs parameter space as discussed in \ref{sec:experimental constraints}. Throughout the paper, the notation $H^\pm \rightarrow \tau^\pm \nu$ is used to denote both $H^+\rightarrow \tau^+ \nu$ and $H^- \rightarrow \tau^- \bar{\nu}$ (similarly for tb channel). For the charged Higgs production cross-section, $\sigma_{H^\pm}$ denotes the sum of $\sigma_{H^+}$ and $\sigma_{H^-}$. Comparison of exclusion limits on charged Higgs parameter space from 13 TeV and 8 TeV CMS results are presented in \ref{sec:experimental constraints} along with the indirect flavour physics constraint coming from $B\rightarrow X_s \gamma$. As mentioned before, the bosonic decay of charged Higgs into $W^\pm h$, $W^\pm H$ and $W^\pm A$ are highly suppressed due to alignment limit and limited phase space. But once the bosonic decay channel is open, the bounds coming from the $H^\pm \rightarrow \tau^\pm \nu$ and $H^+ \rightarrow t\bar{b}$ become weak \cite{Kling:2015uba,Coleppa:2014cca}. In this paper, the latest result from CMS collaboration \cite{Sirunyan:2019zdq} is used in Type I scenario in the mass range $M_{H^\pm}\in[100,160]$ GeV where the mass splitting $M_{H^\pm}-M_A=85$ GeV and $M_{H^\pm}\sim M_H$ is still allowed by $T$ parameter constraint. 
\label{sec:intro}
\label{sec:intro}

\section{Two Higgs Doublet Model (2HDM) Review }
For 2HDM, the most general scalar potential \cite{Branco:2011iw} is 
\begin{eqnarray}
\mathcal{V}(\Phi_1,\Phi_2)&=&m_{11}^2\Phi_1^{\dagger}\Phi_1 + m_{22}^2\Phi_2^{\dagger}\Phi_2 - [m^2_{12}\Phi_1^\dagger \Phi_2 +h.c.]  \nonumber \\ 
&+&\frac{1}{2}\lambda_1(\Phi_1^\dagger \Phi_1)^2 + \frac{1}{2}\lambda_2(\Phi_2^\dagger \Phi_2)^2 + \lambda_3(\Phi_1^\dagger \Phi_1)(\Phi_2^\dagger\Phi_2) \nonumber \\ 
&+&\lambda_4 (\Phi_1^\dagger \Phi_2)(\Phi_2^\dagger\Phi_1) +[\frac{\lambda_5}{2}(\Phi_1^\dagger \Phi_2)^2 + h.c. ]
\label{potential}
\end{eqnarray} 
where $\Phi_{1,2}$ are two isospin doublets with hypercharge $Y=1/2$. To avoid tree level FCNC a $\mathcal{Z}_2$ symmetry is imposed under which $\Phi_1\rightarrow\Phi_1$ and $\Phi_2\rightarrow-\Phi_2$. This symmetry is softly broken by
  the parameter $m_{12}\neq 0$. The parameters $m_{12}$ and $\lambda_5$ are considered real assuming $CP$ invariance. The two Higgs doublets are parameterized as
\begin{eqnarray}
\Phi_i =  \left(\begin{array}{c}\phi^+_i\\
 \frac{v_i +\rho_i+i\eta_i}{\sqrt{2}} \end{array}\right)
\end{eqnarray}
where
$\langle \rho_1 \rangle=v_1$, $\langle \rho_2 \rangle=v_2$ , $\tan\beta=v_2/v_1$ and $v=\sqrt{v_1^2 + v_2^2}\approx246$
GeV.

The physical mass eigenstates are given by
\begin{eqnarray}
\left(\begin{array}{c}
G^\pm\\
H^\pm\\
\end{array}\right)&=&R(\beta)\left(\begin{array}{c}
\phi_1^\pm\\
\phi_2^\pm\\
\end{array}\right), \nonumber \\
\left(\begin{array}{c}
G\\
A\\
\end{array}\right)&=&R(\beta)\left(\begin{array}{c}
\eta_1\\
\eta_2\\
\end{array}\right),\nonumber \\
\left(\begin{array}{c}
H\\
h\\
\end{array}\right)&=&R(\alpha)\left(\begin{array}{c}
\rho_1\\
\rho_2\\
\end{array}\right)
\end{eqnarray} 
Here $G^{\pm}$ and $G$ are the Nambu-Goldstone bosons that
are eaten as the longitudinal components of the massive
gauge bosons. The rotation matrix is given by 

\begin{eqnarray}
R(\theta)=
\left(\begin{array}{cc}
\cos\theta & \sin\theta\\
-\sin\theta & \cos\theta\\
\end{array}\right) 
\end{eqnarray}

Minimization of the scalar potential in Eq.[\ref{potential}] gives
\begin{eqnarray}
m_{11}^2&=&m_{12}^2\tan\beta -\frac{v_1^2 \lambda_1}{2} - \frac{v_2^2 \lambda_{345}}{2} \nonumber\\
m_{22}^2&=&m_{12}^2\tan^{-1}\beta -\frac{v_2^2 \lambda_2}{2} - \frac{v_1^2 \lambda_{345}}{2}
\end{eqnarray}
where $\lambda_{345} = \lambda_3 + \lambda_4 + \lambda_5$.
 The parameters $\lambda_{i}$ written in terms of the physical parameters $M_{H^\pm}, M_A, M_H, M_h, m_{12}$ and the mixing angles $\alpha$ and $\beta$ $(\tan\beta=v_2/v_1)$ are 
\begin{eqnarray}
\lambda_1 &=& \frac{M_{H}^2 R(\alpha)^2_{11}v_1+M_{h}^2 R(\alpha)^2_{12}v_1-m_{12}^2v_2}{v_1^3} \nonumber\\
\lambda_2 &=& \frac{M_{H}^2 R(\alpha)^2_{21}v_2+M_{h}^2 R(\alpha)^2_{22}v_2-m_{12}^2v_1}{v_2^3} \nonumber\\
\lambda_3 &=& \frac{M_H^2 R(\alpha)_{11}R(\alpha)_{21}+M_h^2 R(\alpha)_{12}R(\alpha)_{22}}{v_1 v_2} \nonumber \\
&+&\frac{m_{12}^2 -v_1 v_2\lambda_{45}}{v_1 v_2} \nonumber \\
\lambda_4 &=&\frac{m_{12}^2}{v_1 v_2}+\frac{M_A^2}{v^2}-\frac{2M_{H^\pm}^2}{v^2} \nonumber \\
\lambda_5 &=&\frac{m_{12}^2}{v_1 v_2}-\frac{M_A^2}{v^2}
\end{eqnarray}
 where $\lambda_{45}=\lambda_4 + \lambda_5$.

The most general Yukawa Lagrangian under the $\mathcal{Z}_2$ symmetry is
\begin{eqnarray}
\mathcal{L}^{\text{2HDM}}_{\text{Yukawa}}&=&-\bar{Q}_LY_u\tilde{\Phi}_u u_R-\bar{Q}_LY_d\Phi_d d_R \nonumber \\
&-&\bar{L}Y_l \Phi_l l_R + h.c.
\end{eqnarray}
where $\Phi_f (f=u,d$ or $l)$ is either $\Phi_1$ or $\Phi_2$ depending on the Yukawa models of 2HDM. The four possible $\mathcal{Z}_2$ charge assignments of the quarks and charged leptons can be summarized in Table \ref{Table1}.

\begin{table}[h!]
  \centering
  \begin{tabular}{|l|l|l|l|l|l|l|l|l|l|l|l|l|}
    \hline
    Model & $\Phi_1$ & $\Phi_2$ & $u_R$ & $d_R$ & $l_R$ & $Q_L,L_L$\\
    \hline
    Type I &$+$&$-$&$-$&$-$&$-$&$+$ \\
    \hline
    TypeII &$+$&$-$&$-$&$+$&$+$&$+$ \\
    \hline
    Type X & $+$ & $-$ & $-$ & $-$  & $+$  & $+$  \\
    \hline
    Type Y & $+$ & $-$ & $-$ & $+$ &  $-$ &  $+$  \\
    \hline
  \end{tabular}
  \caption{Charge assignment under $\mathcal{Z}_2$ symmetry to avoid FCNC at tree level.}
  \label{Table1}
\end{table} 

In Type I 2HDM, the second Higgs doublet $\Phi_2$ couples to the fermions, so all the quarks and charged leptons get their masses from the VEV of $\Phi_2$ (ie. $v_2$). In Type II 2HDM, up-type quarks couple to $\Phi_2$ whereas down-type quarks and charged leptons couple to $\Phi_1$. Hence in Type II up-type quarks get masses from $v_2$ and down-type quarks and charged leptons get masses from $v_1$. The Higgs sector of Minimal Supersymmetric Standard Model (MSSM) is a special 2HDM whose Yukawa interaction is of Type II. For Type X (also called Lepton Specific Model), the quark sector is similar to Type I but the charged leptons are coupled to $\Phi_1$ and finally in the Type Y (also called Flipped Model) the quark sector is similar to Type II but the leptons are coupled to $\Phi_2$. Among them, Type II 2HDM has been most widely investigated because of its resemblance with MSSM.  

The Yukawa interactions of $H^\pm$ with quarks and leptons take the form
\begin{eqnarray}
\mathcal{L}_Y&=&-\frac{\sqrt{2}}{v}H^+\bar{u}[\xi_d V M_d P_R - \xi_u M_u V P_L]d \nonumber \\
&-& \frac{\sqrt{2}}{v} H^+ \xi_l\bar{\nu}M_l P_R l
\end{eqnarray}
where $V$ is the CKM matrix and $P_{R,L}=\frac{1}{2}(1\pm\gamma_5)$ are the chirality projection operators.
\begin{table}[h!]
  \centering
  
  \begin{tabular}{|l|l|l|l|l|l|l|l|l|l|l|l|l|}
    \hline
    Model & $\xi_d$ & $\xi_u$ & $\xi_l$\\
    \hline
    Type I & $\cot\beta$ & $\cot\beta$ & $\cot\beta$\\
    \hline
    TypeII & $-\tan\beta$ & $\cot\beta$ & $-\tan\beta$\\
    \hline
    Type X & $\cot\beta$ & $\cot\beta$ & $-\tan\beta$\\
    \hline
    Type Y & $-\tan\beta$ & $\cot\beta$ & $\cot\beta$\\
    \hline
  \end{tabular}
  \caption{Choices of the couplings $\xi_f$ for the four Yukawa models of 2HDM.}
\end{table}

\label{sec:2HDM review}

\section{$H^{\pm}$ production and decay Channels}

\begin{figure}[htp]
\centering
 \includegraphics[width=0.25\textwidth]{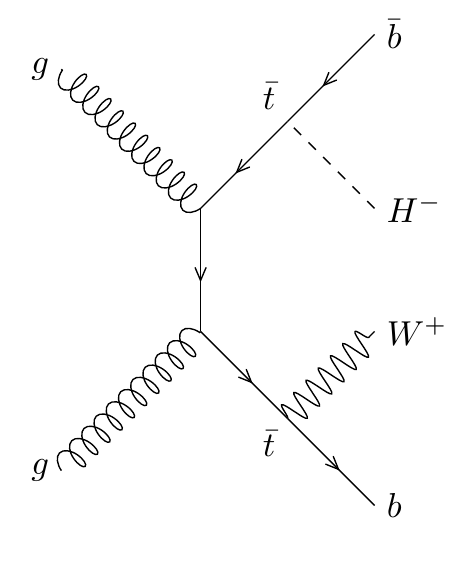}
 \includegraphics[width=0.25\textwidth]{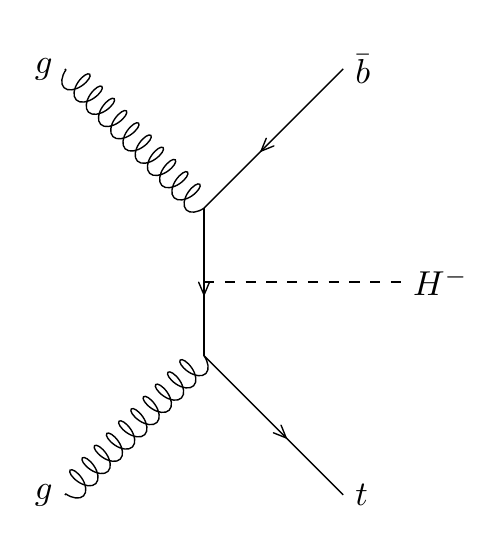}
  \includegraphics[width=0.25\textwidth]{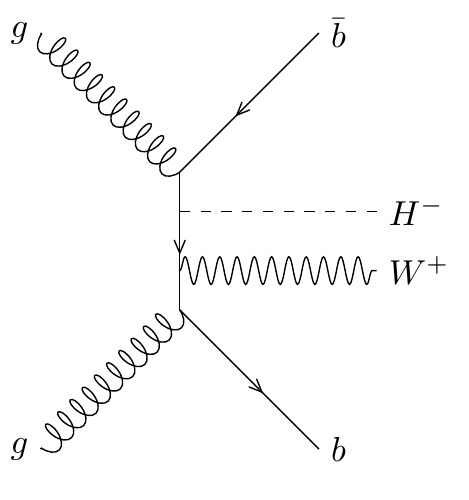}
 \caption{ Leading order (LO) diagrams for charged Higgs are shown. The double resonant top pair production (top diagram) is the dominant process for light $H^\pm.$ The single resonant top quark production (middle diagram) is dominant for heavy $H^\pm$. In the intermediate $H^\pm$ scenario ($M_{H^\pm}\sim M_t$), both of these 
production channels along with the non-resonant top quark production (bottom diagram) are taken into account. }
 \label{sigma}
\end{figure}

The production cross section of charged Higgs depends on its mass with respect to top quark and can be classified into three categories. The light charged Higgs scenario is defined where the mass of charged Higgs is light enough ($M_{H^\pm}\lesssim 150$ GeV) such that the on-shell decay of top quark, $t\rightarrow H^+b$ is allowed. The production cross section for light scenario is simply given by the product of top pair production (double-resonant mode) $p p \rightarrow t \bar{t}$ times the branching fraction of top into charged Higgs $t\rightarrow H^+b$. The $p p \rightarrow t \bar{t}$ cross section has been computed at NNLO in QCD including resumation of NNLL soft gluon terms using the code {\sc Top}$++2.0$ \cite{Czakon:2011xx}. Heavy charged Higgs scenario is defined for $M_{H^\pm}\gtrsim200$ GeV where the charged Higgs mass is sufficiently large compared to top quark. In this scenario, the dominant charged Higgs production channel is the associated production with a top quark (single-resonant mode) $pp \rightarrow tbH^{\pm}$. The production cross section for the heavy charged Higgs boson computed
in the 4FS and 5FS schemes in Refs.\cite{Flechl:2014wfa,Degrande:2015vpa}, and combined together to obtain the total cross section using the Santander matching scheme \cite{Harlander:2011aa} for different values of $\tan\beta$\footnote{Since the charged Higgs production cross sections scales with the glue-glue luminosity, in the mass range of $200-600$ GeV, the production cross section increases by a factor of $4-6$ from 8 TeV to 13 TeV. }. The intermediate charged Higgs scenario is considered for $M_{H^\pm}$ close to top quark i.e. $150\lesssim M_{H^\pm}\lesssim200$ GeV. In this region, the non-resonant top quark production mode also contributes along with the single-resonant and double-resonant modes.  Cross sections at NLO QCD accuracy in the 4FS scheme as given in Ref.\cite{Degrande:2016hyf} are considered. Fig.\ref{sigma} shows the leading order (LO) diagrams of charged Higgs production in the three scenarios\footnote{The cross sections are provided in \url{https://twiki.cern.ch/twiki/bin/view/LHCPhysics/LHCHXSWGMSSMCharged} }. Since the $H^\pm$ interaction to the quark sector of Type I and Type X and similarly for Type II and Type Y are same, the production cross sections in different models are related by  $\sigma^{H^\pm}_{\text{Type \hspace{.2mm}I}}=\sigma^{H^\pm}_{\text{Type \hspace{.2mm}X}}$ and $\sigma^{H^\pm}_{\text{Type \hspace{.2mm}II}}=\sigma^{H^\pm}_{\text{Type \hspace{.2mm}Y}}$. 

\begin{figure}[htp]
\centering
 \includegraphics[width=0.35\textwidth]{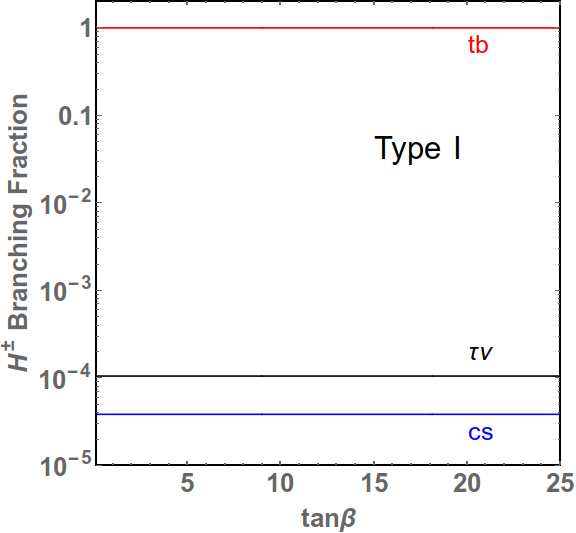}
 \includegraphics[width=0.35\textwidth]{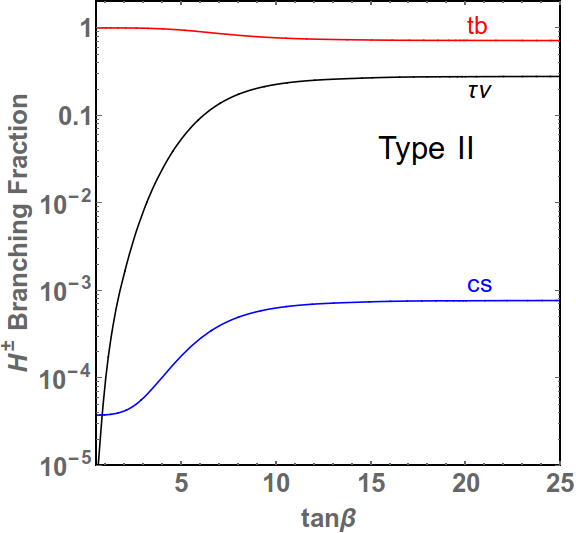}
  \includegraphics[width=0.35\textwidth]{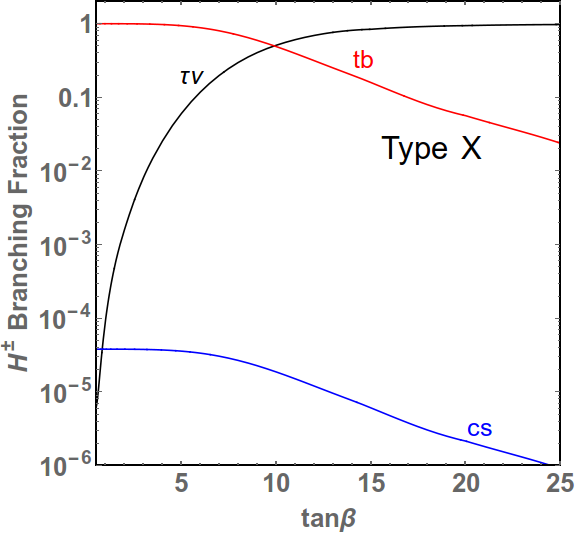}
 \includegraphics[width=0.35\textwidth]{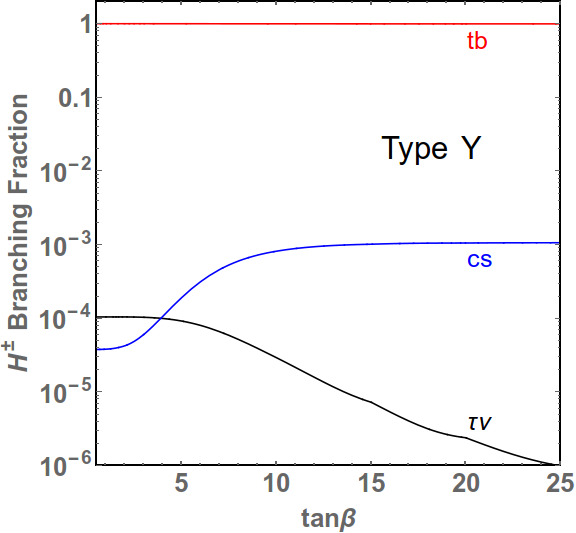}
 \caption{Branching fractions of charged Higgs into the dominant fermionic sectors as a function of $\tan\beta$ for $M_{H^\pm} = 250$ GeV. The alignment limit $\sin(\beta - \alpha)\rightarrow 1$ and degenerate $M_{H^\pm}$, $M_H$ and $M_A$ are considered to prevent $H^\pm \rightarrow W^\pm \phi$ ($\phi=h,H,A$) and satisfy $T$ parameter constraint.  }
 \label{branching fraction}
\end{figure}

For the case of charged Higgs fermionic decays, in Type I all the fermionic couplings are proportional to $\cot\beta$ and hence the branching fractions are independent of $\tan\beta$. The $\tau \nu$ channel is the dominant decay channel for light charged Higgs in Type I. However for heavy charged Higgs scenario in Type I, the Br($H^\pm \rightarrow \tau \nu$) is suppressed by $M_{\tau}^2/M_{t}^2$ over Br($H^+ \rightarrow t\bar{b}$), leading to nearly $100\%$ branching fraction in $t\bar{b}$ channel . In Type II and Type X the lepton sector coupling to $H^\pm$ being proportional to $\tan\beta$, the decay into $\tau \nu$ is dominant for light $H^\pm$ and quite sizable for heavy $H^\pm$ for $\tan\beta \gtrsim 1$. As seen in Fig.\ref{branching fraction} for heavy $H^\pm$ scenario in Type X, the $H^\pm$ branching fraction to $\tau \nu$ starts dominating over $t\bar{b}$ channel for large $\tan\beta$. In Type Y, because of the $\cot\beta$ dependence in the lepton sector the $\tau \nu$ channel gets suppressed compared to the hadronic decay modes (dominantly into $t\bar{b}$ for heavy $H^\pm$). The branching fractions computed using the public code {\sc Hdecay} \cite{Djouadi:1997yw,Djouadi:2018xqq} are shown in Fig.\ref{branching fraction} for $M_{H^\pm}=250$ GeV, for all Yukawa types of 2HDM. The code {\sc Hdecay} also includes the three-body decay of charged Higgs ie. $H^+ \rightarrow t^* \bar{b} \rightarrow W^+ b \bar{b}$ below the two-body decay threshold of $H^+ \rightarrow t \bar{b}$ mode \cite{Djouadi:1995gv}. Note that the branching fraction of $H^\pm$ into the fermionic sector is given for situations where there are no $H^\pm$ decay into the neutral scalars. 

Apart from the fermionic decays, $H^\pm$ can also decay to $W^\pm$ and neutral scalars $h,H$ or $A$. The couplings to $W^\pm$ and neutral scalars are (all fields are incoming)

\begin{eqnarray}
H^\mp W^\pm h &:& \frac{\mp i g}{2} \cos(\beta - \alpha)(p_\mu - p^\mp_\mu) \nonumber \\
H^\mp W^\pm H &:& \frac{\mp i g}{2} \sin(\beta - \alpha)(p_\mu - p^\mp_\mu) \nonumber \\
H^\mp W^\pm A &:& \frac{g}{2}(p_\mu - p^\mp_\mu)
\end{eqnarray}
 where $p_\mu$ and $p^\mp_\mu$ are the momenta of the neutral and charged scalars. In the alignment limit $\sin(\beta-\alpha)\rightarrow 1$ (which is considered throughout the paper) $H^\pm$ decay to $h$ is suppressed. The decay into the $H$ and $A$ channels depend on the mass splitting allowed by $T$ parameter. In the generic 2HDM, there are no mass relations between $H^\pm, H$ and $A$ unlike MSSM and for some parameter choice, the bosonic decays can be more dominant compared to the fermionic decays once the channels are open. 
 
\label{sec:production scoss section}

\section{Experimental Constraints} 

The theoretical constraints of 2HDM consist of vacuum stability \cite{PhysRevD.18.2574,Nie:1998yn}, perturbative unitarity \cite{PhysRevD.16.1519,Grinstein:2015rtl} and tree level unitarity \cite{Kanemura:1993hm,Akeroyd:2000wc,Arhrib:2000is}. The Electro-Weak Precision Observables (EWPOs) $S (0.05\pm0.11), T (0.09\pm0.13) $ and $U (0.01\pm0.11)$ \cite{Grimus:2007if,Grimus:2008nb}, specially the $T$ parameter \cite{PhysRevD.98.030001} restrict the mass splitting of $H^\pm$, $H$ and $A$. In this paper, $M_{H^\pm}=M_H=M_A$ is considered to impose the exclusion limits from the $H^\pm \rightarrow \tau^\pm \nu$ and $H^+ \rightarrow t \bar{b}$ channels over the mass range $M_{H^\pm} \in [80,2000]$ GeV. Perturbative unitarity for a wide region of $\tan\beta$ can be satisfied by proper choice of the 
 soft $\mathcal{Z}_2$ breaking parameter, $m_{12}^2=M_A^2 \sin\beta\cos\beta$. The theoretical constraints are checked using the package {\sc 2Hdmc-1.7.0} \cite{Eriksson:2009ws}. Alignment limit $\sin(\beta-\alpha)\rightarrow 1$ is the most favoured condition by the experimentalists. In this limit the couplings of the neutral scalar $h$ in 2HDM is similar to SM Higgs boson and can be identified as the observed 125 GeV Higgs boson. In the alignment limit the other $CP$ even scalar, $H$ behaves as gauge-phobic i.e. its coupling to the gauge bosons $W^\pm /Z$ is very suppressed. In the context of charged Higgs analysis for $H^\pm \rightarrow \tau^\pm \nu$ and $H^+ \rightarrow t\bar{b}$ channels, the alignment limit is useful as it completely suppresses the $H^\pm \rightarrow W^\pm h$ decay. LEP experiments \cite{Abbiendi:2013hk} have given limits on the mass of charged Higgs in 2HDM from the charged Higgs searches in Drell Yan events $e^+ e^- \rightarrow Z/\gamma \rightarrow H^+ H^-$, excluding $M_{H^\pm} \lesssim 80$ GeV (Type II) and $M_{H^\pm} \lesssim 72.5$ GeV (Type I) at $95\%$ confidence level. Among the constraints from $B$ meson decays (flavour physics constraints), the $B\rightarrow X_s\gamma$ decay \cite{Amhis:2014hma} puts a very strong constraint on Type II and Type Y 2HDM, excluding $M_{H^\pm} \lesssim 580$ GeV and almost independently of $\tan\beta$. For Type I and Type X, $B\rightarrow X_s\gamma$ constraint is sensitive only for low $\tan\beta.$ So for $M_H^{\pm} \lesssim 580$ GeV, Type II and Type Y are not considered further. 
 
\begin{figure}{}
\centering
 \includegraphics[width=0.35\textwidth]{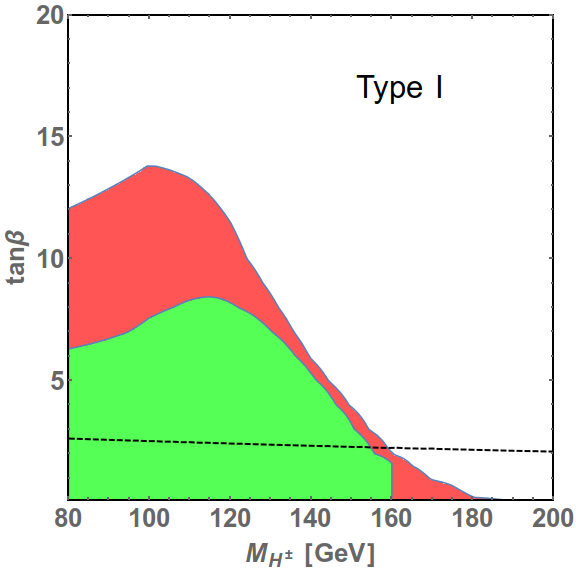}
 \includegraphics[width=0.35\textwidth]{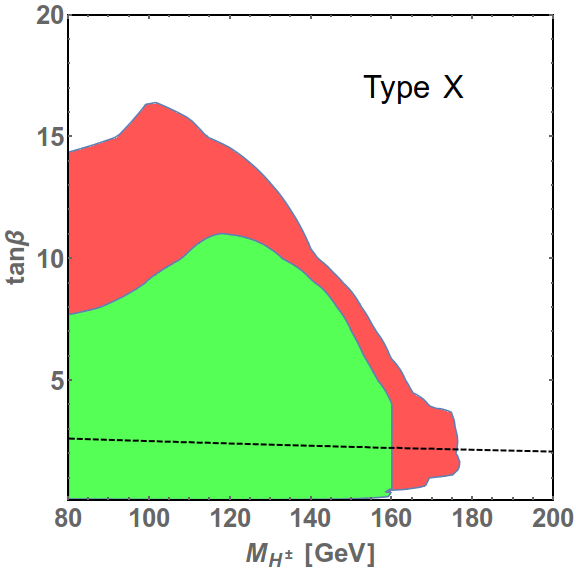}
 \caption{Exclusion region in Type I and Type X from the upper limits on $\sigma_{H^\pm}BR(H^\pm \rightarrow \tau^\pm \nu)$ CMS 13 TeV observations are shown in red colour. The green colour shows the exclusion region from the upper limits on BR$(t \rightarrow H^+ b )$BR($H^\pm \rightarrow \tau^\pm \nu$) CMS 8 TeV observations. The region below the black dashed line is excluded from BR($B\rightarrow X_s\gamma$) constraint.} 
 \label{taunu}
\end{figure}

The LHC experiments have already set limits on the $M_{H^\pm}-\tan\beta$ plane using $\sqrt{s}=8$ TeV observations from $H^\pm \rightarrow \tau^\pm \nu$ \cite{Aad:2014kga,Khachatryan:2015qxa} and $H^+ \rightarrow t \bar{b}$ \cite{Khachatryan:2015qxa,Aad:2015typ} channels. For $M_{H^\pm}\in[80-160]$ GeV, the most important constraint comes from the $H^\pm \rightarrow \tau^\pm\nu$ channel\footnote{The constraints of charged Higgs decaying in the fermionic sector is useful only when the charged Higgs bosonic decays are suppressed.}. The exclusion regions are shown with green colors in Fig.\ref{taunu} using the 8 TeV CMS results at an integrated luminosity of 19.7 fb$^{-1}$  for Type I and Type X. In Type X the leptonic coupling being proportional to $\tan\beta$ excludes a slightly larger region of $\tan\beta$. Using the upper bounds on the $\sigma_{H^\pm}$BR$(H^\pm \rightarrow \tau^\pm \nu)$ from the latest CMS results \cite{Sirunyan:2019hkq} for $\sqrt{s}=13$ TeV at an integrated luminosity of 35.9 fb$^{-1}$, a much larger region of $\tan\beta$ is excluded as shown in red colours in Fig. \ref{taunu} for both Type I and Type X . Just above $M_{H^\pm}=160$ GeV $\tan\beta \lesssim 1$ is allowed by this channel in Type X model. This is because the exclusion in Type X at low $\tan\beta$ is less severe compared to Type I.
 
\begin{figure}{}
\centering
 \includegraphics[width=0.36\textwidth]{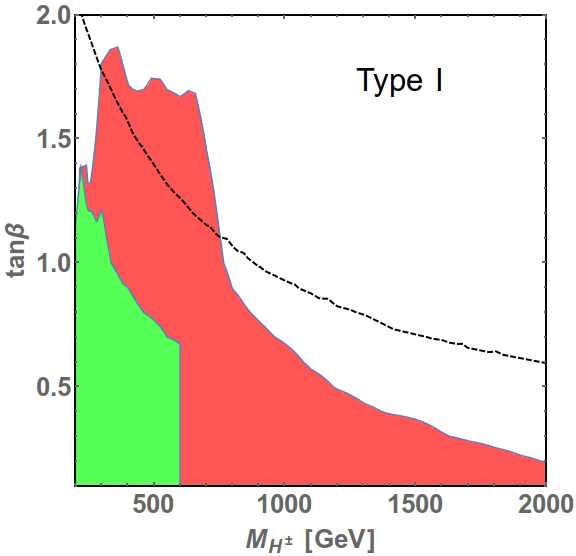}
 \includegraphics[width=0.36\textwidth]{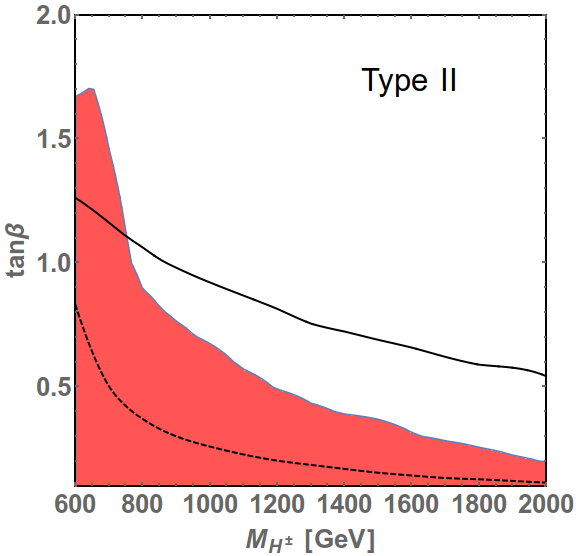}
 \caption{Exclusion region in Type I and Type II from the upper limits on $\sigma_{H^\pm}BR(H^+ \rightarrow t\bar{b})$ CMS 13 TeV observations are shown in red colour. The green colour (upper plot) shows the exclusion region from the upper limits on $\sigma(pp\rightarrow \bar{t}(b)H^+)$ CMS 8 TeV observations assuming BR($H^+ \rightarrow t \bar{b})=1$. The region below the black dashed line is excluded by BR($B\rightarrow X_s\gamma$) constraint and the region below the continuous black line in Type II (bottom plot) is excluded by BR$(B_s \rightarrow \mu^+ \mu^-)$ constraint.}
 \label{tb}
\end{figure}

 For $M_{H^\pm}$ greater than top quark, the constraint coming from the $\tau \nu$ channel does not put any significant bound in the $M_{H^\pm} - \tan\beta$ parameter space. Therefore in the higher mass range the  $H^+ \rightarrow t\bar{b}$ channel has to be studied. The $tb$ channel, unlike the $\tau \nu$ channel, is not clean enough and suffers from various QCD backgrounds, but sophisticated analysis are used to study the $tb$ channel in both 8 TeV and 13 TeV by the CMS collaboration. The CMS 8 TeV upper limit on $\sigma(pp \rightarrow t(b)H^+)$ assuming BR($H^+ \rightarrow t \bar{b}$) = 100$\%$ \cite{Khachatryan:2015qxa} restricts the parameter space for $200<M_{H^\pm}<600$ GeV. The exclusion region using the 8 TeV results are shown in green colours in Fig.\ref{tb} for Type I. Recent paper from CMS collaboration \cite{CMS:1900zym} for $\sqrt{s}=13$ TeV and 35.9 fb$^{-1}$ puts an upper limit at $95\%$ CL on $\sigma_{H^\pm}\text{BR}(H^+ \rightarrow t\bar{b})$ with the single-lepton and dilepton final states combined. The resulting exclusion region in the $M_{H^\pm}-\tan\beta$ plane for heavy charged Higgs $M_{H^\pm}\in [200,2000]$ GeV in Type I  and $M_{H^\pm}\in [600,2000]$ in Type II are shown in Fig.\ref{tb} with red colours. Since the charged Higgs leptonic decay mode in TypeY is much suppressed compared to $H^+ \rightarrow t\bar{b}$ mode and for Type X scenario, the $H^+ \rightarrow t \bar{b}$ mode is dominant for low $\tan\beta$ as shown in Fig.\ref{branching fraction} (bottom two plots). The exclusion regions of Type X and Type Y are equilvalent to the exclusion regions of Type I and Type II respectively.  

\begin{figure}[htp]
\centering
\includegraphics[width=0.35\textwidth]{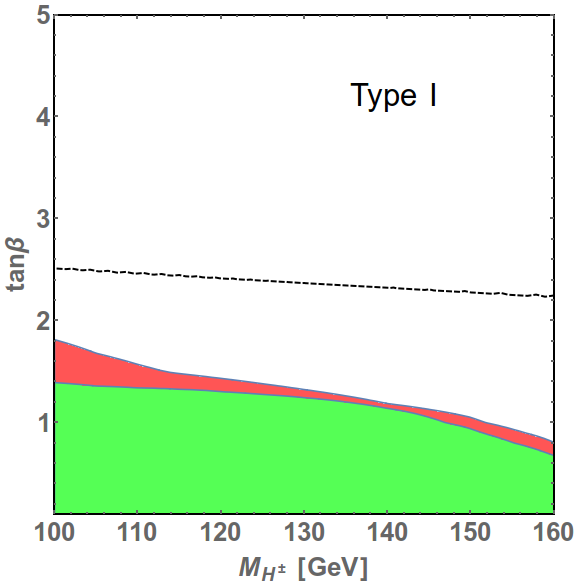}
 \includegraphics[width=0.35\textwidth]{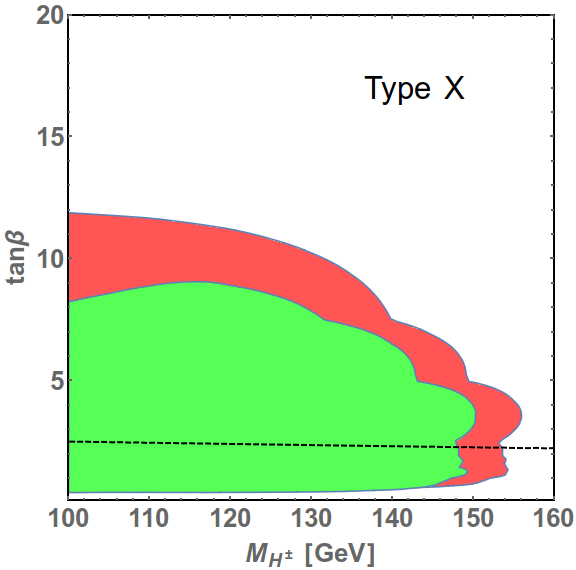}
 \caption{Exclusion region in Type I and Type X from the upper limits on $\sigma_{H^\pm}BR(H^\pm \rightarrow \tau^\pm \nu)$ CMS 13 TeV observations are shown in red colour. The green colour shows the exclusion region from the upper limits on BR$(t \rightarrow H^+ b )$BR($H^\pm \rightarrow \tau^\pm \nu$) CMS 8 TeV observations. The region below the black dashed line is excluded from BR($B\rightarrow X_s\gamma$) constraint. Here the mass difference $M_{H^\pm}-M_A=85$ GeV for $M_{H^\pm}\in[100,160]$ GeV and $M_{H^\pm}\sim M_{H}$ is considered. The exclusion region is less compared to Fig.\ref{taunu} where the masses of $H^\pm$, $H$ and $A$ are nearly same. }
 \label{taunu_2nd}
\end{figure}

  So far, the charged Higgs decay to the gauge boson and neutral scalars $H^\pm \rightarrow W^\pm H/A$ are not considered by assuming nearly mass degeneracy of $H^\pm, H$ and $A$. But once the bosonic decays are kinematically allowed, charged Higgs can significantly decay into these channels. Fig.\ref{taunu_2nd} shows the exclusion regions coming from the $H^\pm \rightarrow \tau^\pm \nu$ channel where the mass difference $M_{H^\pm}-M_A=85$ GeV is considered for $M_{H^\pm}\in[100-160]$ GeV and $M_{H^\pm}\sim M_H$. The red regions are excluded by using the upper limits on $\sigma_{H^\pm}BR(H^\pm \rightarrow \tau^\pm \nu)$ CMS 13 TeV observations and the green regions are excluded by using the upper limits on BR$(t \rightarrow H^+ b )$BR($H^\pm \rightarrow \tau^\pm \nu$) CMS 8 TeV observations. For this choice of mass difference, the exclusion regions are less compared to Fig.\ref{taunu} because of significant decay of $H^\pm$ into $W^\pm A$. As mentioned above, in Type X the leptonic coupling being proportional to $\tan\beta$ excludes a larger region compared to Type I.  
   
\begin{figure}[htp]
\centering
\includegraphics[width=0.35\textwidth]{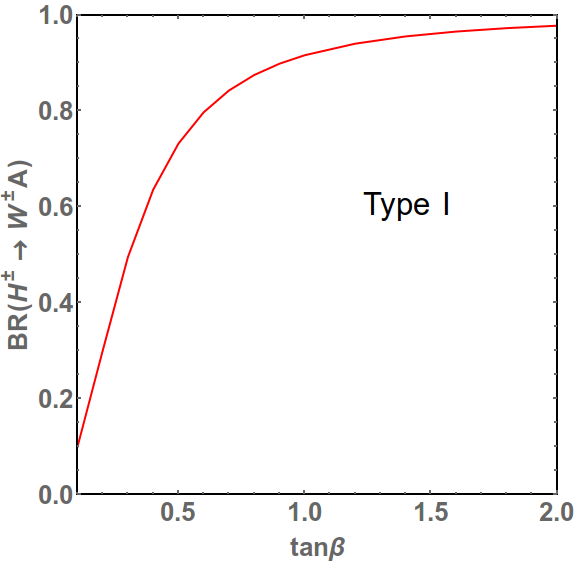}
 \includegraphics[width=0.35\textwidth]{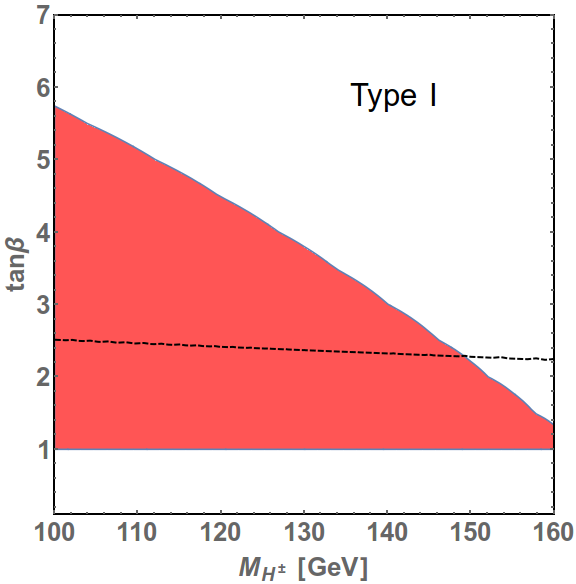}
 \caption{Branching fraction of $H^\pm \rightarrow W^\pm A$ as a function of $\tan\beta$ (upper plot) for $M_{H^\pm} - M_A=85$ GeV and $M_{H^\pm}=150$ GeV. Exclusion region (bottom plot) in Type I is shown in red colour from the first upper limits on BR$(t \rightarrow H^+ b$) CMS 13 TeV observations in the charged Higgs decay mode: $H^\pm \rightarrow W^\pm A$ and $A \rightarrow \mu^+ \mu^-$ with the assumptions BR$(H^\pm \rightarrow W^\pm A)=1$ and BR$(A\rightarrow \mu^+\mu^-)=3\times10^{-4}$. For the charged Higgs bosonic decay channel the mass difference $M_{H^\pm}-M_A=85$ GeV for $M_{H^\pm}\in[100,160]$ GeV is considered. The region below the black dashed line is excluded by BR($B\rightarrow X_s\gamma$) constraint.  }
 \label{WA}
\end{figure}

  The CMS collaboration \cite{Sirunyan:2019zdq} recently studied the scenario where the mass difference of $H^\pm$ and $A$ is $\sim 85$ GeV for $M_{H^\pm}\in[100-160]$ GeV. The charged Higgs produced in $p p $ collision in LHC at an integrated luminosity of 35.9 fb$^{-1}$, decays dominantly into $W^\pm$ and $A$ with final states $e\mu\mu$ or $\mu\mu\mu$. In this analysis, the CMS assumed BR($H^\pm \rightarrow W^\pm A) = 1$ and BR($A\rightarrow \mu^+ \mu^-) = 3\times 10^{-4}$. Also this is the first experimental result in the channel $H^\pm \rightarrow W^\pm A$, $A\rightarrow \mu^+ \mu^-$ at LHC to put upper limits on BR$(t \rightarrow H^+ b)$. Such low branching fraction of $A\rightarrow \mu^+ \mu^-$ can be realized in Type I 2HDM where BR$(A\rightarrow \mu^+ \mu^-) \sim2.4\times 10^{-4}$ for $A\in[15,75]$ GeV and it goes very well with the CMS assumption. The other assumption, BR($H^\pm \rightarrow W^\pm A)=1$, is satisfied in Type I scenario when $\tan\beta \geq 1$ as seen in Fig.\ref{WA}. In Type I scenario the charged Higgs coupling to the fermionic sector being proportional to $\cot\beta$, the assumption  BR($H^\pm \rightarrow W^\pm A)=1$, starts to fail for $\tan\beta < 1$. The theoretical constraints can be satisfied with proper choice of $m^2_{12}$ and the oblique parameter $T$ can be satisfied by considering $M_{H^\pm}\cong M_H$. The observed upper limit at 95$\%$ CL on BR$(t \rightarrow H^+ b)$ for $M_H^{\pm}\in[100,160]$ GeV and $M_{H^\pm}-M_A=85$ GeV with the above assumptions are used to find the exclusion region. In Fig.\ref{WA} the red region (above $\tan\beta \geq 1$) shows the exclusion region where we have smooth fitted the observed CMS upper limit on BR$(t \rightarrow H^+ b)$ in the range of 0.63 to 2.9$\%$. Other 2HDMs like Type II and Type Y are not considered as for this  mass range of charged Higgs Type II and Type Y are ruled out by $B\rightarrow X_s \gamma$ constraint. Unlike Type I where all the fermionic couplings of $A$ is proportional to $\cot\beta$, in Type X the pseudoscalar coupling to the lepton sector is proportional to $\tan\beta$ whereas its coupling to the quark sector is proportional to $\cot\beta$. Thus the BR$(A\rightarrow \mu^+\mu^-)$ increases with $\tan\beta$. The CMS assumption is satisfied in Type X scenario only when $\tan\beta$ is close to $1$ and for this situation the theoretically estimated BR$(t\rightarrow H^+ b)$ is same as in Type I (because of the same coupling) and above the upper limit of the CMS observation. Comparing Figs.\ref{taunu_2nd} and \ref{WA}, the exclusion regions coming from the $\tau\nu$ channel is weak once the $H^\pm \rightarrow W^\pm A$ channel is open. Fig.\ref{WA} (bottom plot) excludes the region of parameter space which are not excluded in Fig.\ref{taunu_2nd} (top plot).

  For completeness, the indirect constraints from the flavour physics is also considered in the paper as the  $B$ meson decay depends strongly on the parameters $M_{H^\pm}$ and $\tan\beta$. The public code {\sc SuperIso-3.7} \cite{Mahmoudi:2008tp} is used for flavour physics computation. As mentioned above, for Type II and Type Y, charged Higgs lighter than $\sim 580$ GeV is completely ruled out for large region of $\tan\beta$ from BR($B\rightarrow X_s \gamma$) constraint \cite{Misiak:2017bgg} which is measured to be $(3.32\pm0.15)\times10^{-4}$ \cite{Amhis:2016xyh}. For Type I (and similarly for Type X) the BR($B\rightarrow X_s \gamma$) constraint excludes $\tan\beta \lesssim 2 $. In Figs.\ref{taunu}, \ref{tb}, \ref{taunu_2nd} and \ref{WA} the region below the black dashed lines are excluded from BR($B\rightarrow X_s \gamma$) observation. In Type II (and similarly in Type Y) for $M_{H^\pm}$ above 600 GeV, the rare decay of $B_s \rightarrow \mu^+ \mu^-$ (the branching fraction of which is measured to be ($3.0\pm0.6\pm0.25)\times10^{-9}$) as reported by LHCb collaboration excludes a greater region of parameter space compared to $B\rightarrow X_s\gamma$ constraint. For the Type II scenario in Fig.\ref{tb} the region below the black continuous line is excluded by BR($B_s \rightarrow \mu^+ \mu^-$) constraint. 
  \hspace{5mm} 
\label{sec:experimental constraints}

\section{Summary and Conclusions}
 The 2HDM is the simplest extension of SM containing charged Higgs. The two most dominant channels, $H^\pm \rightarrow \tau^\pm \nu$ and $H^+ \rightarrow t \bar{b}$, for the search of $H^\pm$ are studied using the latest CMS results for 
$\sqrt{s}=13$ TeV at an integrated luminosity of 35.9 fb $^{-1}$. The $\tau \nu $ channel excludes a large region of $\tan\beta < \mathcal{O}(15)$ for charged Higgs mass less than 160 GeV both in Type I and Type X. For heavy charged Higgs, the $\tau \nu$ channel does not lead to any significant
 constraint on the parameter space. However, in this case, 
the $tb$ channel excludes a significant range of values of $\tan\beta$ 
in Type I and II and the same behaviour is carried over to Type X and Type Y. The exclusion regions obtained from the 13 TeV CMS results are compared with the exclusion regions from 8 TeV CMS results. Exclusion bounds from $B$ meson decays are also discussed for all Yukawa types of 2HDM. The fermionic channels are studied for situations where the exotic decays of charged Higgs into gauge boson and neutral scalars ($H^\pm \rightarrow W^\pm/h/H/A$) are suppressed either by alignment limit or due to limited phase space. But once the bosonic decay channels are open, they can be the dominant charged Higgs decay channels and the constraints from $H^\pm \rightarrow \tau^\pm \nu$ and $H^+ \rightarrow t\bar{b}$ will be less restrictive. The CMS collaboration for the first time studied the exotic bosonic decay channel $H^\pm \rightarrow W^\pm A$ and $A \rightarrow \mu^+ \mu^-$ to put upper limits on BR($t \rightarrow H^+ b$) for $M_{H^\pm} \in [100,160]$ GeV with a mass splitting of $M_{H^\pm}-M_A=85$ GeV. These results are used to exclude a significant parameter space of charged Higgs in Type I 2HDM which is not excluded by the $\tau \nu$ channel. It is expected that a significant parameter space of charged Higgs will be excluded in all Yukawa types of 2HDM (as well as in MSSM) if these exotic bosonic decay channels of charged Higgs are analysed by CMS or ATLAS collaborations for various charged Higgs mass ranges.

\section{Acknowledgments}
\noindent
The author would like to thank Ravindra K. Verma and Aravind H. Vijay for some useful discussions. The author also acknowledges Pankaj Jain for discussions
and useful comments on the paper.

\bibliographystyle{JHEP}
\bibliography{Reference.bib}

\end{document}